\documentclass[aps,pre,twocolumn,groupedaddress,showpacs,showkeys]{revtex4}

\usepackage{graphicx}
\usepackage{epsfig}

\begin{document}

\title[Supersymmetry and Models for Interacting Particles]
      {Supersymmetry and Models for Two Kinds of Interacting
       Particles}

\author{Thomas Guhr\dag\ and Heiner Kohler\ddag}
\email{thomas.guhr@matfys.lth.se}
\email{kohler@tphys.uni-heidelberg.de}
\affiliation{\dag\
         Matematisk Fysik, LTH, Lunds Universitet,
         Box 118, 22100 Lund, Sweden\\
         \ddag\
         Institut f\"ur theoretische Physik, Universit\"at
         Heidelberg, Philosophenweg 19, Heidelberg, Germany}

\date{\today}

\begin{abstract}
We show that Calogero--Sutherland models for interacting particles
have a natural supersymmetric extension. For the construction, we use
Jacobians which appear in certain superspaces. Some of the resulting
Hamiltonians have a direct physics interpretation as models for two
kinds of interacting particles. One model may serve to describe
interacting electrons in a lower and upper band of a
quasi--one--dimensional semiconductor, another model corresponds to
two kinds of particles confined to two different spatial directions
with an interaction involving tensor forces.
\end{abstract}



\maketitle

Calogero~\cite{cal69,cal71} and Sutherland~\cite{sut72} introduced
models for interacting particles, which have much in common with a
group theoretical construction by Dyson~\cite{DYS2}. Various
modifications of these models have been studied, see the detailed
reviews in Refs.~\cite{OP,CAL} and references therein. A typical model of
the Calogero--Sutherland type is defined by the Schr\"odinger equation
for $N$ particles in one dimension at positions $x_n, \ n=1,\ldots,N$,
\begin{equation}
H\Psi_N^{(\beta)}(x,\kappa) =   
  \left(\sum_{n=1}^N\kappa_n^2\right)\Psi_N^{(\beta)}(x,\kappa) \ ,
\label{csse}
\end{equation}
with the Hamiltonian
\begin{eqnarray}
H = -\sum_{n=1}^N\frac{\partial^2}{\partial x_n^2} +
\beta\left(\frac{\beta}{2}-1\right) \sum_{n<m}\frac{1}{(x_n-x_m)^2} \ .
\label{calo}
\end{eqnarray}
The many particle wavefunction $\Psi_N^{(\beta)}(x,\kappa)$ depends on
$N$ quantum numbers $\kappa_n, \ n=1,\ldots,N$ whose squares add up to
the energy on the right hand side of Eq.~(\ref{csse}).  The
Hamiltonian~(\ref{calo}) consists of a kinetic and a distance
dependent interaction term. The strength is measured by the parameter
$\beta$, the interaction vanishes for $\beta=0$ and $\beta=2$.
Usually, one adds $N$ confining potentials to the
Hamiltonian~(\ref{calo}).  This renders the system a bound state
problem. Apart from this, they do not significantly affect the
structure of the model. Thus, we will not work with confining
potentials in the sequel.

In this contribution, we present a most natural extension of these
Calogero--Sutherland models. In our construction, we employ
superspaces. We will arrive at various new models which are likely to
be exactly solvable. Calogero--Sutherland models were related to
supersymmetric quantum mechanics in Refs.~\cite{susy1,susy2}. Our
approach and the ensuing models are different from this. Our models
also extend a recent supersymmetric construction~\cite{ser,serves}.
Importantly, our models allow for a physics interpretation which is
also most natural. One of the models describes a
quasi--one--dimensional problem, the other one a two--dimensional
one. One has previously tried to generalize the models of the
type~(\ref{csse}) to higher space
dimensions~\cite{gos97,kah98}. Again, our construction and the results
are different from this.

To prepare for the derivation of the new models in superspace, we
briefly sketch the group theoretical connection for the models in
ordinary space. With the Vandermonde determinant $\Delta_N(x) =
\prod_{n<m}(x_n-x_m)$, the Jacobian on the space of the
real--symmetric $N\times N$,\ the Hermitean $N\times N$ and the
quaternion self--dual $2N\times 2N$ matrices labeled with the
parameter $\beta=1,2,4$, respectively, takes the form
$|\Delta_N(x)|^\beta$. These spaces are non--compact forms of the
symmetric spaces ${\rm U}(N)/{\rm O}(N)$, ${\rm U}(N)/1$ and ${\rm
U}(2N)/{\rm Sp}(2N)$. The corresponding radial Laplace--Beltrami
operator reads
\begin{equation}
\Delta_x = \sum_{n=1}^N \frac{1}{|\Delta_N(x)|^\beta}
\frac{\partial}{\partial x_n}|\Delta_N(x)|^\beta
\frac{\partial}{\partial x_n} \ .
\label{opo}
\end{equation}
The equation $\Delta_x \Phi_N^{(\beta)}(x,\kappa) =
-\left(\sum_{n=1}^N\kappa_n^2\right) \Phi_N^{(\beta)}(x,\kappa)$ for
the eigenfunction $\Phi_N^{(\beta)}(x,\kappa)$ is mapped onto the $N$
particle Schr\"odinger equation~(\ref{csse}) with the ansatz
$\Phi_N^{(\beta)}(x,k) = \Psi_N^{(\beta)}(x,k)/
\Delta_N^{\beta/2}(x)\Delta_N^{\beta/2}(k)$.  The parameter $\beta$ is
now viewed as positive and continuous, such that the coordinates $x_n,
\ n=1,\ldots,N$ span a space more general than the (radial parts of)
the symmetric spaces.

For the supersymmetric generalization, we consider the two sets of
$k_1$ variables $s_{p1}, \ p=1,\ldots,k_1$ and of $k_2$ variables
$s_{p2}, \ p=1,\ldots,k_2$ and the function
\begin{equation}
B_{k_1k_2}(s) = \frac{\prod_{p<q}(s_{p1}-s_{q1})^{\beta_1}
                      \prod_{p<q}(s_{p2}-s_{q2})^{\beta_2}}
                     {\prod_{p,q}(s_{p1}-cs_{q2})^{\sqrt{\beta_1\beta_2}}} \ ,
\label{cb1}
\end{equation}
with the parameters $\beta_1,\beta_2\geq 0$ and $c=\pm i$.  For
certain values of these parameters, Eq.~(\ref{cb1}) is the Jacobian
(or Berezinian)~\cite{BER} on symmetric superspaces. In the case
$\beta_1=\beta_2=2$ and $c=+i$, the function~(\ref{cb1}) is the
Jacobian on Hermitean supermatrices, i.e.~on the non--compact form of
${\rm U}(k_1/k_2)/1$~\cite{EFE83,TG}, where ${\rm U}(k_1/k_2)$ is the
unitary supergroup~\cite{BER,KAC1,KAC2}. Apart from an absolute value
sign which is unimportant here, the choices $\beta_1=1$, $\beta_2=4$,
$c=+i$ and $\beta_1=4$, $\beta_2=1$, $c=-i$ in Eq.~(\ref{cb1}) yield
the Jacobians~\cite{EFE83,GUH4} for the two forms of the symmetric
superspace ${\rm Gl}(k_1/2k_2)/{\rm OSp}(k_1/2k_2)$, respectively.
They are denoted ${\rm AI|AII}$ and ${\rm AII|AI}$ in the
classification of Ref.~\cite{MRZ1}.  Here, ${\rm Gl}(k_1/2k_2)$ is the
general linear supergroup and ${\rm OSp}(k_1/2k_2)$ is the
orthosymplectic supergroup~\cite{BER,KAC1,KAC2}.  The imaginary unit
in the parameter $c$ stems from a Wick--type--of rotation of the
variables $s_{p2}$, which was performed for a convergence
reason~\cite{EFE83}.  Although not needed here, we keep it for now to
make the notation compatible with the literature.  The
function~(\ref{cb1}) induces the operator
\begin{eqnarray}
\Delta_s &=&\frac{1}{\sqrt{\beta_1}}
     \sum_{p=1}^{k_1}
     \frac{1}{B_{k_1k_2}(s)}
           \frac{\partial}{\partial s_{p1}}
           B_{k_1k_2}(s)
           \frac{\partial}{\partial s_{p1}} \nonumber\\ 
          && \quad + \frac{1}{\sqrt{\beta_2}}\sum_{p=1}^{k_2}
           \frac{1}{B_{k_1k_2}(s)}
           \frac{\partial}{\partial s_{p2}}
           B_{k_1k_2}(s)
           \frac{\partial}{\partial s_{p2}} \ .
\label{cb2}
\end{eqnarray}
The prefactors $1/\sqrt{\beta_1}$ and $1/\sqrt{\beta_2}$ in front of
the sums are such that $\Delta_s$ becomes the radial Laplace--Beltrami
operator on the three symmetric superspaces mentioned above for the
corresponding choices of the parameters $\beta_1$, $\beta_2$ and
$c$. However, we emphasize that arbitrary positive values for
$\beta_1$ and $\beta_2$ will be considered in the sequel, while the
parameter $c$ remains restricted to $c=\pm i$. We map the eigenvalue
equation for the operator $\Delta_s$ with eigenfunctions
$\varphi_{k_1k_2}^{(c,\beta_1,\beta_2)}(s,r)$ onto the equation
\begin{eqnarray}
&&\widetilde{H}\psi_{k_1k_2}^{(c,\beta_1,\beta_2)}(s,r) = \nonumber\\
&&\qquad \left(\sum_{p=1}^{k_1}\frac{r_{p1}^2}{\sqrt{\beta_1}}+
    \sum_{p=1}^{k_2}\frac{r_{p2}^2}{\sqrt{\beta_2}}\right)
           \psi_{k_1k_2}^{(c,\beta_1,\beta_2)}(s,r) \ ,
\label{cb4}
\end{eqnarray}
where the wavefunctions are related by the ansatz
$\varphi_{k_1k_2}^{(c,\beta_1,\beta_2)}(s,r)=
\psi_{k_1k_2}^{(c,\beta_1,\beta_2)}(s,r)/(B_{k_1k_2}(s)
B_{k_1k_2}(r))^{1/2}$. The resulting operator reads
\begin{eqnarray}
\widetilde{H}&=&
 -\frac{1}{\sqrt{\beta_1}}\sum_{p=1}^{k_1}
  \frac{\partial^2}{\partial s_{p1}^2}
 -\frac{1}{\sqrt{\beta_2}}\sum_{p=1}^{k_2}
  \frac{\partial^2}{\partial s_{p2}^2} \nonumber\\
&& \qquad +\sum_{p<q}\frac{g_{11}}{\left(s_{p1}-s_{q1}\right)^2} 
          +\sum_{p<q}\frac{g_{22}}{\left(s_{p2}-s_{q2}\right)^2} \nonumber\\
&& \qquad -\sum_{p,q}\frac{g_{12}}{\left(s_{p1}-cs_{q2}\right)^2} 
\label{cb5}
\end{eqnarray}
with constants
\begin{eqnarray}
g_{jj}&=&\sqrt{\beta_j}\left(\frac{\beta_j}{2}-1\right)  \ , \quad j=1,2 \ , 
                                             \nonumber\\
g_{12}&=&\frac{1}{2}\left(\sqrt{\beta_1}-\sqrt{\beta_2}\right)
\left(\frac{1}{2}\sqrt{\beta_1\beta_2}+1\right) \ .
\label{con}
\end{eqnarray}
The constant on the right hand side of Eq.~(\ref{cb4}) is interpreted
as energy later on. We write it in terms of two sets of variables
$r_{p1}, \ p=1,\ldots,k_1$ and $r_{p2}, \ p=1,\ldots,k_2$ to which we
refer as quantum numbers. It turns out convenient to take the scaling
factors $1/\sqrt{\beta_1}$ and $1/\sqrt{\beta_2}$ into the definition.
In Ref.~\cite{serves}, a similar construction was performed. However,
the resulting operator depends on one parameter only. The natural
dependence on two strength parameters $\beta_1$ and $\beta_2$ is an
essential point in the present study. Thus, our approach contains the
one in Ref.~\cite{serves} as a special case.

The family of models~(\ref{cb5}) ought to be exactly solvable. For the
parameter values corresponding to the symmetric superspaces, the
wavefunctions can be written as supergroup
integrals~\cite{TG,GGT,GUKO2}. Thus, the wavefunctions are
uniquely characterized by the two sets of quantum numbers $r_{p1}, \
p=1,\ldots,k_1$ and $r_{p2}, \ p=1,\ldots,k_2$.  This feature should
extend to all values $\beta_1,\beta_2\ge 0$, since the operators
$\Delta_s$ and $\widetilde{H}$ are analytic in the upper right
quadrant of the complex plane $\beta_1+i\beta_2$. However, a final
statement needs more mathematical work.

We present a physical interpretation. The operators $\Delta_s$ and
$\widetilde{H}$ are not Hermitean, which is solely due to the
Wick--type--of rotation mentioned above. We consider $c=+i$ and undo
this rotation by replacing $is_{p2}$ with $s_{p2}$. By also
introducing the momenta $\pi_{p1}=-i\partial/\partial s_{p1}$ and
$\pi_{p2}=-i\partial/\partial s_{p2}$, we transform the
operator~(\ref{cb5}) into the Hermitean Hamiltonian for two kinds of
$k_1$ particles at positions $s_{p1}, \ p=1,\ldots,k_1$ and $k_2$
particles at positions $s_{p2}, \ p=1,\ldots,k_2$ on the $s$ axis,
\begin{eqnarray}
H&=&
  \sum_{p=1}^{k_1} \frac{\pi_{p1}^2}{2m_1}
 +\sum_{p=1}^{k_2} \frac{\pi_{p2}^2}{2m_2}
 +\sum_{p<q}\frac{g_{11}}{\left(s_{p1}-s_{q1}\right)^2}\nonumber\\
 && \quad -\sum_{p<q}\frac{g_{22}}{\left(s_{p2}-s_{q2}\right)^2}
           -\sum_{p,q}\frac{g_{12}}{(s_{p1}-s_{q2})^2}\ ,
\label{PI1}
\end{eqnarray}
with now canonical conjugate variables satisfying
$[\pi_{pj},s_{ql}]=\delta_{pq}\delta_{jl}$. We notice that the mass
$m_1=\sqrt{\beta_1/4}$ is positive, while the mass
$m_2=-\sqrt{\beta_2/4}$ is negative. The interactions are according to
Eq.~(\ref{con}) repulsive or attractive, depending on the choices for
$\beta_1$ and $\beta_2$.  The particles of the same kind interact,
this interaction vanishes for $\beta_1=2$ or $\beta_2=2$. Two
particles of different kind also interact.  This interaction vanishes
for $\beta_1=\beta_2$ and the problem decouples into two known
models~(\ref{calo}). For $k_1=0$ or $k_2=0$, we recover the
models~(\ref{calo}).

\begin{figure}
\begin{center} 
\epsfig{figure=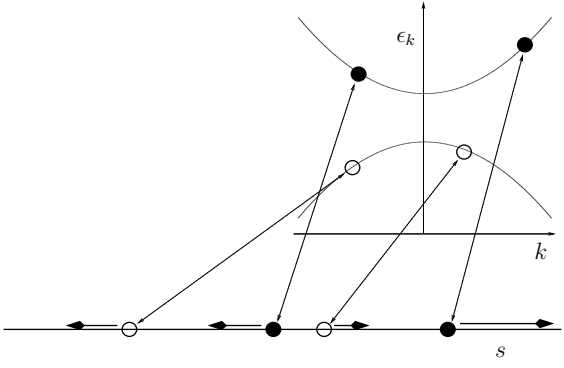,width=7.5cm,angle=0}
\caption{\label{fig1} 
Electrons in the upper (black circles) and lower
(open circles) band of a quasi--one--dimensional semiconductor. The
dispersion relations $\epsilon_k$ as function of the wave number $k$
are indicated by the parabola and the inverted parabola. The particles
are then mapped onto the $s$ axis in the bottom part.}
\end{center}
\end{figure}

The Hamiltonian~(\ref{PI1}) may also serve to model the motion of
electrons in a quasi--one--dimensional semiconductor, see
Fig.~\ref{fig1}. The particles at positions $s_{p1}$ with positive
mass
$m_1$ are identified with the electrons subject to a periodic
potential in the upper band close to the gap. The electrons in the
lower band have a dispersion relation $\epsilon_k$ as function of the
wave number $k$ whose second derivative, i.e.~the inverse mass, is
negative~\cite{KIT}.  They are identified with the particles at
positions $s_{p2}$ that have negative mass $m_2$.

\begin{figure}
\begin{center} 
\epsfig{figure=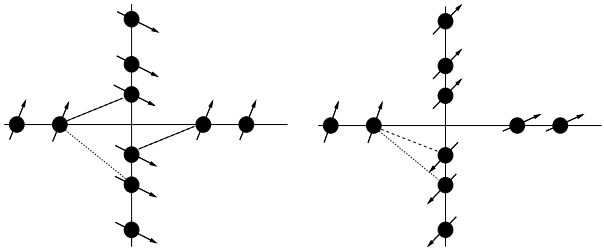,width=8cm,angle=0}
\caption{\label{fig2} Two realizations of the two--dimensional
model. Left: $2k_1=4$ particles on the $s_1$ axis and $2k_2=6$ particles
on the $s_2$ axis. The dipole vectors on the same axis have the same
direction. Tensor forces are indicated as thicker and thinner dashed
lines, corresponding to the strength of the force. Right: a case with
different directions of the dipole vectors on different sides of the
same axis.}
\end{center}
\end{figure}

The models~(\ref{calo}) are based on the ordinary unitary group and on
associated symmetric spaces. There are other models in ordinary space
related to the ordinary orthogonal and symplectic groups~\cite{OP}.
They have a natural supersymmetric extension as well.  We introduce
the two sets of $2k_1$ variables $s_{p1}, \ p=1,\ldots,k_1$ and of
$2k_2$ variables $s_{p2}, \ p=1,\ldots,2k_2$. We notice that the
number of variables in the second set has to be even. Instead of
$2k_1$, we could also consider an odd number $2k_1+1$. As the ensuing
models differ only slightly, we restrict ourselves to $2k_1$. We study
the function
\begin{eqnarray}
&& C_{2k_12k_2}(s) = \nonumber\\
&& \frac{\prod_{p<q}(s_{p1}^2-s_{q1}^2)^{\beta_1}
                       \prod_{p<q}(s_{p2}^2-s_{q2}^2)^{\beta_2}
                       \prod_{p=1}^{k_2}s_{p2}^{\beta_2}}
                      {\prod_{p,q}(s_{p1}^2+s_{q2}^2)^{\sqrt{\beta_1\beta_2}}} 
                             \ , \qquad 
\label{bke}
\end{eqnarray}
which is the Jacobian (or Berezinian) on the supergroup ${\rm
OSp}(2k_1/2k_2)$ for $\beta_1=\beta_2=2$, see a derivation in
Ref.~\cite{HCUOSP}. For $\beta_1=1$, $\beta_2=4$ and $\beta_1=4$,
$\beta_2=1$, Eq.~(\ref{bke}) gives the Jacobians on the two forms of
the symmetric superspace ${\rm OSp}(2k_1/2k_2)/{\rm Gl}(k_1/k_2)$
which are named ${\rm CI|DIII}$ and ${\rm DIII|CI}$ in
Ref.~\cite{MRZ1}.  We proceed exactly as before and derive an
eigenvalue equation of the form~(\ref{cb4}) where the operator now
reads
\begin{eqnarray}
\widetilde{H}&=&
  -\frac{1}{\sqrt{\beta_1}}\sum_{p=1}^{k_1}
   \frac{\partial^2}{\partial s_{p1}^2}
  -\frac{1}{\sqrt{\beta_2}}\sum_{p=1}^{k_2}
   \frac{\partial^2}{\partial s_{p2}^2} \nonumber\\
&& \quad -g_{11}\sum_{p<q}\frac{2s_{p1}^2+2s_{q1}^2}
                                {\left(s_{p1}^2-s_{q1}^2\right)^2}\nonumber\\
&& \quad +g_{22}\left(\sum_{p<q}\frac{2s_{p2}^2+2s_{q2}^2}
                                      {\left(s_{p2}^2-s_{q2}^2\right)^2}
              + \sum_{p=1}^{k_2}\frac{1}{2s_{p2}^2}\right)\nonumber\\
&& \quad -g_{12} \sum_{p,q}\frac{2s_{p1}^2-2s_{q2}^2}
                                  {\left(s_{p1}^2+s_{q2}^2\right)^2}
          +\sum_{p,q}\frac{2h_{12}}{s_{p1}^2+s_{q2}^2} \ ,
\label{cb5osp}
\end{eqnarray}
with $g_{11}$, $g_{22}$ and $g_{12}$ as given in Eq.~(\ref{con}) and
with
\begin{equation}
h_{12}=\frac{1}{4}\sqrt{\beta_1\beta_2}\left(\sqrt{\beta_1}-\sqrt{\beta_2}\right) \ .
\label{hcon}
\end{equation}
The operator~(\ref{cb5osp}) remains invariant when replacing any of
the variables by its negative. Due to this symmetry, $\widetilde{H}$
itself is an Hermitean operator and can be viewed as a Hamiltonian,
there is no need to undo the Wick--type--of rotation. We now come to a
physical interpretation. In a straightforward calculation, the
Hamiltonian $H=2\widetilde{H}$ can be cast into the form
\begin{eqnarray}
H&=&
  \sum_{p=1}^{2k_1}\frac{\pi_{p1}^2}{2m_1}
 +\sum_{p=1}^{2k_2}\frac{\pi_{p2}^2}{2m_2}
 +\sum_{p=1}^{2k_1}\frac{f_1}{s_{p1}^2}
 +\sum_{p=1}^{2k_2}\frac{f_2}{s_{p2}^2} \nonumber\\ 
&& \quad +\sum_{p<q}\frac{h_{11}}{\left(s_{p1}-s_{q1}\right)^2}
         +\sum_{p<q}\frac{h_{22}}{\left(s_{p2}-s_{q2}\right)^2} \nonumber\\
&& \quad -\sum_{p,q}\frac{h_{12}}{s_{p1}^2+s_{q2}^2} \nonumber\\
&& \quad +\sum_{p,q}\frac{\left(\vec{e}_{pq}\cdot\vec{\sigma}_1\right)
                          \left(\vec{e}_{pq}\cdot\vec{\sigma}_2\right)
                                       -\vec{\sigma}_1\cdot\vec{\sigma}_2/2}
                         {s_{p1}^2+s_{q2}^2} \ .
\label{PI2}
\end{eqnarray}
We exploit the symmetry by writing $H$ as a Hamiltonian for $2k_1$
plus $2k_2$ particles, i.e.~$k_1$ plus $k_2$ pairs of particles
sitting at positions $(-s_{p1},+s_{p1})$ on the $s_1$ axis and
$(-s_{p2},+s_{p2})$ on the $s_2$ axis. This is a two--dimensional
situation. Here, we assume that the initial condition is invariant
under mirror reflection of the positions $s_{p1}$ and $s_{p2}$.  Each
particle on the $s_j$ axis carries a (fixed and non--quantized) dipole
vector $\vec{\sigma}_j=\sigma (\cos\vartheta_j,\sin\vartheta_j), \
j=1,2$. The masses $m_j=\sqrt{\beta_j/4} , \ j=1,2$ are both
positive. The interaction comprises three parts. First, there are
central forces with strengths
\begin{equation}
f_1 = +\frac{\beta_1}{8}\left(\frac{\beta_1}{2}-1\right) \quad {\rm and} \quad
f_2 = -\frac{\beta_2}{8}\left(\frac{\beta_2}{2}-1\right) \ .
\label{coup2orth}
\end{equation}
Second, there are distance dependent forces between the particles
on the same axis with strengths
\begin{equation}
h_{jj} = \sqrt{\beta_j}\left(\frac{\beta_j}{2}-1\right)
            +\sigma^2\cos 2\vartheta_j \ , \quad j=1,2
\label{couporth}
\end{equation} 
and there is a distance dependent force between the particles on
different axes with strength $h_{12}$ as given by Eq.~(\ref{hcon}).
Third, there are tensor or dipole--dipole forces. The last term of the
Hamiltonian~(\ref{PI2}) is a two--dimensional dipole--dipole
interaction~\cite{BOMO1} between the particles on different axes. The
unit vector $\vec{e}_{pq}$ points from the particle $p$ on the $s_1$
axis to the particle $q$ on the $s_2$ axis.  The strength of the
dipoles follows from the relation $\sigma^2
\cos(\vartheta_1+\vartheta_2) = 2g_{12}$ with $g_{12}$ as defined in
Eq.~(\ref{con}).  Consistently, the tensor force also acts between the
dipoles on the same axis. As the latter are parallel, the tensor force
acquires a form identical to the distance dependent interaction. This
explains the additional term in the constants $h_{jj}, \ j=1,2$ as
compared to the constants $g_{jj}$. In Fig.~\ref{fig2} we
illustrate the model. Other choices of the parameters are also
possible, leading, for example, to different angles of the dipoles on
the positive and negative side of the same axis, see Fig.~\ref{fig2}.

Again, the family of models~(\ref{PI2}) should be exactly
solvable. For the parameter values corresponding to ${\rm
OSp}(2k_1/2k_2)$, the wavefunctions are supergroup
integrals~\cite{HCUOSP}. Analytical continuation in $\beta_1+i\beta_2$
should be possible. A rigorous proof has yet to be given.

To gain some first intuition for the solutions of these models, we
consider the simplest case $k_1=k_2=1$ and $c=+i$ of Eq.~(\ref{cb4}).
The operator~(\ref{cb5}) has then a simple structure which yields the
wavefunction
\begin{eqnarray}
&&\psi_{11}^{(+i,\beta_1,\beta_2)}(s,r) = \nonumber\\
&& \quad \exp\left(\pm\frac{i\left(\sqrt{\beta_1}s_{11}-i\sqrt{\beta_2}s_{12}\right)
                         \left(r_{11}-ir_{12}\right)}
                       {\sqrt{\beta_1}-\sqrt{\beta_2}}\right)\nonumber\\
 && \quad \frac{z^\nu {\bf H}^{\mp}_{\nu}(z)}
            {((s_{11}-is_{12})(r_{11}-ir_{12}))^{\sqrt{\beta_1\beta_2}/2}} \ .
\label{solu11}
\end{eqnarray}
Here, ${\bf H}^{\mp}_{\nu}(z)$ is the Hankel function~\cite{ABR} of
order $\nu=\sqrt{\beta_1\beta_2}/2+1/2$ and we use the dimensionless,
complex variable $z=(\sqrt{\beta_2}
r_{11}-i\sqrt{\beta_1}r_{12})(s_{11}-is_{12})/(\sqrt{\beta_2}-\sqrt{\beta_1})$. The
appearance of the differences in the denominator in Eq.~(\ref{solu11})
is typical for superspaces, see Refs.~\cite{GUKO2}. In ordinary
space, those differences are found in the numerator, see
Refs.~\cite{GUKOP1,GUKO1}. In particular, this affects the behavior of
the wavefunctions at the origin. Work on further analytical results is
in progress.

In conclusion, we derived natural supersymmetric extensions of models
for interacting particles. The corresponding physics is most natural
as well, involving two kinds of particles. We presented two possible
applications, a quasi--one--dimensional and a two--dimensional
one. From a more general perspective, one might say that our results
yield a conceptually new interpretation of supersymmetry. In
high--energy physics, the physical bosons and fermions are represented
by commuting and anticommuting variables, see Ref.~\cite{PER}. This is
also so in the interacting boson--fermion model~\cite{IAIS} of nuclear
physics. In chaotic and disordered systems~\cite{EFE83}, the commuting
and anticommuting variables serve to considerably reduce the numbers
of the degrees of freedom, they are not seen as physical
particles. Here, we showed that the radial coordinates on certain
superspaces can be viewed as the positions of interacting particles.

This work was finished at the Centro Internacional de Ciencias (CIC)
in Cuernavaca, Mexico. We thank Director Thomas Seligman for
hospitality.  TG acknowledges financial support form the Swedish
Research Council.


\end{document}